\documentclass[aps,prd,twocolumn,showpacs,groupedaddress,nofootinbib,superscriptaddress,10pt]{revtex4-1} 

\usepackage[T1]{fontenc}
\usepackage{lmodern}

\usepackage[dvipsnames,x11names]{xcolor}

\usepackage[colorlinks=true,linkcolor=Blue,citecolor=ForestGreen,urlcolor=NavyBlue]{hyperref}

\usepackage{graphicx}
\usepackage[sort&compress]{natbib}
\usepackage{lipsum}
\usepackage{morefloats}
\usepackage[pdf]{pstricks}
\usepackage{subfigure}
\usepackage{amsmath}
\usepackage{amssymb}
\usepackage{amsfonts}
\usepackage{rotating}
\usepackage{cancel}
\usepackage{mathtools}
\usepackage{bbm}
\usepackage{dsfont}
\usepackage{bbold}
\usepackage{multirow}
\usepackage{ulem}
\usepackage{physics}
\usepackage{orcidlink}

\newcommand{\qmax}{\Lambda}

\newcommand{\IFIC}{Instituto de F\'{i}sica Corpuscular, Centro Mixto Universidad de Valencia-CSIC, Institutos de Investigaci\'{o}n de Paterna, Aptdo. 22085, E-46071 Valencia, Spain}
\newcommand{\DepFTUV}{Departamento de F\'isica Te\'orica}
\newcommand{\INFNCatania}{Istituto Nazionale di Fisica Nucleare, Sezione di Catania, Dipartimento di Fisica ``Ettore Majorana'', Universit\`a di Catania, Via Santa Sofia 64, I-95123 Catania, Italy}
\begin{document}

\frenchspacing

\title{\boldmath Inverse problem in femtoscopic correlation functions: The $T_{cc}(3875)^+$ state}

\author{M. Albaladejo\orcidlink{0000-0001-7340-9235}}
\email{Miguel.Albaladejo@ific.uv.es}
\affiliation{\IFIC}%

\author{A. Feijoo\orcidlink{0000-0002-8580-802X}}%
\email{edfeijoo@ific.uv.es}
\affiliation{\IFIC}%
 
\author{I. Vida\~na\orcidlink{0000-0001-7930-9112}}
\email{isaac.vidana@ct.infn.it}
\affiliation{\INFNCatania}

\author{J. Nieves\orcidlink{0000-0002-2518-4606}}%
\email{Juan.M.Nieves@ific.uv.es}
\affiliation{\IFIC}%

\author{E. Oset\orcidlink{0000-0002-4462-7919}}%
\email{Eulogio.Oset@ific.uv.es}
\affiliation{\DepFTUV\ and \IFIC}%

\begin{abstract}
We study here the inverse problem of starting from the femtoscopic correlation functions of related channels and analyze them with an efficient tool to extract the maximum information possible on the interaction of the components of these channels, and the existence of possible bound states tied to this interaction. The method is flexible enough to accommodate non-molecular components and the effect of missing channels relevant for the interaction. We apply the method to realistic correlation functions for the $D^{*+}D^0$ and $D^{*0}D^+$ channels derived consistently from the properties of the $T_{cc}(3875)^+$ and find that we can extract the existence of a bound state, its nature as a molecular state of the $D^{*+}D^0$ and $D^{*0}D^+$ channels, the probabilities of each channel, as well as scattering lengths and  effective ranges for the channels, together with the size of the source function, all of them with a relatively good precision.
\end{abstract}

\maketitle

\section{Introduction}
The measurement of correlation functions when studying the hadron-pairs production in heavy ion collision is becoming a useful tool to learn about hadron dynamics, as proved in a number of experimental \cite{STAR:2014dcy,ALICE:2017jto,STAR:2018uho,ALICE:2018ysd,ALICE:2019hdt,ALICE:2019eol,ALICE:2019buq,ALICE:2019gcn,ALICE:2020mfd,ALICE:2021szj,ALICE:2021cpv,Fabbietti:2020bfg} and theoretical papers \cite{Morita:2014kza,Ohnishi:2016elb,Morita:2016auo,Hatsuda:2017uxk,Mihaylov:2018rva,Haidenbauer:2018jvl,Morita:2019rph,Kamiya:2019uiw,Kamiya:2021hdb,Liu:2023uly,Albaladejo:2023pzq,Ikeno:2023ojl,Torres-Rincon:2023qll,Liu:2023wfo}. In particular, the ALICE collaboration has already started to work on the $D$ sector with the study of the $Dp$ channel \cite{ALICE:2022enj} and a systematic study of the correlations function for the exotic mesons is planned by the same collaboration for the future \cite{ALICE:2022wwr}.

In a recent paper \cite{Vidana:2023olz} we studied the correlation functions of the $D^{*+}D^0$ and $D^{*0}D^+$ channels in connection with the $T_{cc}(3875)^+$ state discovered in Refs.~\cite{LHCb:2021vvq,LHCb:2021auc}. The study took advantage of using an on shell factorization of the Bethe-Salpeter equation which leads to the formulation of Koonin-Prat \cite{Koonin:1977fh,Pratt:1990zq,Bauer:1992ffu,Liu:2023uly} but in a simplified way. A different method based on coordinate space, rather than momentum space, is used in Ref.~\cite{Kamiya:2022thy} obtaining results in qualitative agreement with Ref.~\cite{Vidana:2023olz}. The purpose of the present paper is to do the inverse problem. This is, starting from the correlation functions for $D^{*+}D^0$ and $D^{*0}D^+$, to employ an analysis tool to obtain the properties of the $T_{cc}(3875)^+$. We present here three methods, applied to this particular case, but which are general and can be used for many problems in hadron physics. The first method is extremely simple and allows one to obtain approximate scattering lengths. The second method uses a $K$-matrix approach that allows us to improve on the first one. Finally, the third method, more sophisticated and consistent with dispersion relations, considers real parts not included in the $K$-matrix approach, and allows us to obtain scattering lengths, effective radii, and the binding and width of the $T_{cc}(3875)^+$ state, this is, all the information concerning the $T_{cc}(3875)^+$ state measured by LHCb collaboration. At the same time, the analysis allows us to determine the molecular compositeness of the the $T_{cc}(3875)^+$ as has been recently shown in Ref.~\cite{Dai:2023cyo} (see also Ref.~\cite{Albaladejo:2022sux}).

\section{Formalism: correlation functions}

We start with a brief discussion about the correlation functions, and how the $D^{\ast}D$ interaction enters into their calculation. We consider the coupled channels $D^{\ast+}D^0$ and $D^{\ast0}D^+$. As discussed in Ref.~\cite{Vidana:2023olz} (where the reader is referred for further details), for the $S$-wave interaction relevant in the limited range of values of the c.m. momentum of the outgoing $D^\ast D$ system, the correlation functions can be written as:
\begin{align}
&C_{D^0D^{*+}}(p_{D^0})=1+4\pi\,\theta(\qmax-p_{D^0}) \int_0^{\infty} drr^2S_{12}(r)\nonumber \\
&\times \Big\{ \left\lvert j_0(p_{D^0}r)+T_{11}(s)\widetilde{G}_{1}(r;s) \right\rvert^2 \\
&+             \left\lvert T_{12}(s)\widetilde{G}_{2}(r;s) \right\rvert^2 - j^2_0(p_{D^0}r) \Big\}\,~, \nonumber 
\label{eq:cf4}
\end{align} 
\begin{align}
&C_{D^+D^{*0}}(p_{D^+})=1+4\pi\,\theta(\qmax-p_{D^+})\int_0^{\infty} drr^2S_{12}(r) \nonumber \\
&\times \Big\{ \left\lvert j_0(p_{D^+}r)+T_{22}(s)\widetilde{G}_{2}(r;s) \right\rvert^2 \\
&+             \left\lvert T_{12}(s)\widetilde{G}_{1}(r;s) \right\rvert^2 - j^2_0(p_{D^+}r) \Big\}\,~, \nonumber 
\label{eq:cf5}
\end{align} 
%
where $T_{ij}(s)$ are the $T$-matrix elements of the coupled channel system, to be further discussed below, and $s \equiv E^2$ is the invariant mass squared of the $D^{\ast}D$ pair. Furthermore, $p_{D^0}$ and $p_{D^+}$ are the c.m. momenta of the $D^0 D^{\ast+}$ and $D^+ D^{\ast0}$ channels, respectively, given by:
\begin{subequations}\label{eq:mom_def}\begin{align}
   p_{D^0} \equiv q & = \frac{\lambda^{1/2}(s,m_{D^0}^2,m_{D^{*+}}^2)}{2\sqrt{s}}         \,,\\
   p_{D^+} \equiv k & = \frac{\lambda^{1/2}(s,m_{D^+}^2,m_{D^{*0}}^2)}{2\sqrt{s}}\,
\end{align}\end{subequations}
being $\lambda(a,b,c)=a^2+b^2+c^2-2ab-2ac-2bc$ the well known K\"{a}llen function. The zeroth order spherical Bessel function has been denoted by $j_0(z)$. 

$S_{12}(r)$ is the so-called source function, assumed to be spherically symmetric and usually parameterized as a gaussian function: 
\begin{equation}\label{eq:source}
S_{12}(r) = \frac{1}{(\sqrt{4\pi})^3 R^3}\mbox{exp}\left(-\frac{r^2}{4R^2}\right)\,, 
\end{equation}
with $R$ a parameter representing the size of the system, typically in the range $1$ to $5\,\text{fm}$, depending on the transverse mass of the initial colliding system, \textit{i.e.}, $pp$ or heavy ions collisions. We shall see that in the inverse problem that we solve, we are also able to get this parameter. 

Finally, the function $\widetilde{G}_i(r;s)$ is given by 
\begin{align}\label{eq:GtildeOriginal}
\widetilde{G}_i(r;s)&=\int_{0}^{\qmax}\!\!\! dq H_i(q) \, \frac{j_0(qr)}{s-\left( \omega_1^{(i)}(q)+\omega_2^{(i)}({q}) \right)^2+i\epsilon} \,,\\
H_i(q) & = \frac{q^2}{2\pi^2} \frac{\omega^{(i)}_1(q)+\omega^{(i)}_2(q)}{2\omega^{(i)}_1(q)\,\omega^{(i)}_2(q)}\,,
\end{align} 
where $\omega^{(i)}_j(q)=\sqrt{m_j^{(i)2}+q^2}$, which for practical purposes is better evaluated using
\begin{align}
\widetilde{G}^{(i)}(r;s)&=\int_{0}^{\qmax} dq\, H_i(q) \frac{j_0(q r)-j_0(p_i r)}{s-\left(\omega_1^{(i)}(q)+\omega_2^{(i)}({q})\right)^2} \nonumber \\
&+ j_0(p_i r)  G_i(s),
\label{eq:Gtilde}
\end{align}
with:
\begin{equation}\label{eq:Gcut}
G_i(s) = \int_{0}^{\qmax} dq\, H_i(q) \frac{1}{s-\left( \omega_1^{(i)}(q) + \omega_2^{(i)}({q}) \right)^2 + i\epsilon}\,.
\end{equation}
The advantage of this form is that, since $\sqrt{s}=\omega^{(i)}_1(p_i)+\omega^{(i)}_2(p_i)$, the first integral in Eq.\ (\ref{eq:Gtilde}) has no longer a singularity, while the second integral is a standard two-point loop function regularized with a cutoff $\Lambda$, that can be evaluated analytically as can be found in Ref.~\cite{Oller:1998hw} (in the erratum).

\section{\boldmath Formalism: $T$-matrix models}

\subsection{Simplified method to obtain the scattering lengths}\label{subsec:ScLe}

We start with a transition potential, which we do not need to specify:
\begin{equation}\label{eq:Vij}
V_{ij} = \left(
           \begin{array}{cc}
             V_{11} & V_{12}   \\[0.1cm]
              V_{21}  &  V_{22}\\
           \end{array}
         \right)\,,
\end{equation}
and the scattering amplitude given by:
\begin{equation}\label{eq:BSeq}
T=[1-VG]^{-1} \, V\,,
\end{equation}
with $G$ a diagonal matrix with elements given by the loop function [Eq.~\eqref{eq:Gcut}] of the intermediate pseudoscalar-vector channels, $G\equiv \text{diag} \left[ G_1(s), G_2(s) \right]$. The value of $\qmax$ in the $G_i(s)$ function indicates the range of the interaction in momentum space \cite{Gamermann:2009uq,Song:2022yvz}.

In this part we ignore the $D^*$ width, which is shown in Ref.~\cite{Dai:2023cyo} to have a very small effect on the scattering lengths. We also ignore in this part the effect of the effective range $r_0$, also shown to be moderate in the study of \cite{Vidana:2023olz}. With the normalization of the $G$ function of Eq.\ (\ref{eq:Gcut}) suited to the field theoretical formulation one has \cite{Dai:2023cyo}
\begin{subequations}\label{eq:Tmat}
\begin{align}
T_{11} & \simeq -\frac{8 \pi \sqrt{s}}{-\frac{1}{a_1}-iq} \,,\\
T_{22} & \simeq -\frac{8 \pi \sqrt{s}}{-\frac{1}{a_2}-ik} \,,
\end{align}
\end{subequations}
where $a_1$, $a_2$ are, respectively, the scattering lengths of the $D^{*+}D^0$ and $D^{*0}D^+$ channels and neglecting the $D^{*}$ width, $a_1$ is real, but $a_2$ is complex because $D^{*0}D^+$ can decay into $D^{*+}D^0$. At the $D^{*0}D^+$ threshold, $k=0$ and the only imaginary part comes from $G_{D^{*+}D^0}$ in Eq.\ (\ref{eq:BSeq}), ${\rm Im}G_{D^{*+}D^0}=-q/8 \pi \sqrt{s}$. Ignoring terms in $q^2$, consistently with also not considering the terms in $r_0q^2/2$ of the effective range expansion, we can write 
\begin{equation}\label{eq:scatl2}
-\frac{1}{a_2}=b_2-ib'_2q \, .
\end{equation}
Let us recall that the c.m. differential cross section in the present normalization is given by \cite{Mandl:1985bg}
\begin{equation}\label{eq:difxsect}
\frac{d\sigma_{i \leftarrow j}}{d\Omega}=\frac{1}{64\pi^2}\frac{1}{s}\frac{p_j}{p_i}|T_{ij}|^2
\end{equation}
with $p_i$, $p_j$ the initial and final momenta respectively, and the optical theorem states \cite{Mandl:1985bg} \
\begin{equation}\label{eq:opticalth}
{\rm Im}T_{22}=-2k\sqrt{s}\sigma_{\rm tot}\,~, \qquad \sigma_{\rm tot}=\sigma_{22}+\sigma_{1 \leftarrow 2} \, .
\end{equation}
Using Eqs.~\eqref{eq:Tmat}, \eqref{eq:scatl2}, \eqref{eq:difxsect}, and \eqref{eq:opticalth}, we immediately find that
\begin{equation}\label{eq:sigma21}
\sigma_{1 \leftarrow 2}=\sigma_{\rm tot}-\sigma_{22}=\frac{1}{k}\frac{ 4 \pi b'_2 q}{b_2^2+(b'_2 q+k)^2} \equiv \frac{1}{64 \pi^2}\frac{4 \pi}{s}\frac{q}{k}|T_{12}|^2.
\end{equation}
We see that by having $T_{11}$ and $T_{22}$ one cannot get $T_{12}$, but we can deduce $\left\lvert T_{12} \right\rvert^2$, which is exactly the magnitude we need for the correlation function. Thus, one can express $|T_{12}|^2$ as
\begin{equation}\label{eq:T21}
|T_{12}|^2=64 \pi^2 s \frac{ b'_2}{b_2^2+(b'_2 q+k)^2} \, .
\end{equation}
Below the second threshold $k$ is purely imaginary ($k \to i\kappa$) and then the denominator of 
Eq.\ (\ref{eq:T21}) becomes $(b_2+\kappa)^2+b^{\prime 2}_2 q^2$.

\subsection{\boldmath $K$-matrix formalism}\label{subsec:KMat}

A quite general parameterization of the $T$-matrix is the $K$-matrix parameterization,

\begin{equation}
T^{-1}(s) = K^{-1}(s) - i \rho(s)~, 
\end{equation}
where $\rho(s)$ is a diagonal matrix with elements 
\begin{equation}
    \rho_{ii}(s) = -\frac{p_i(s)}{8\pi\sqrt{s}} = \text{Im}\,G_i(s)~.
\end{equation}
This parameterization fulfills unitarity as long as the $K$-matrix elements do not have a right-hand cut. Therefore, they can be rational functions, or functions with left-hand cuts, or a combination thereof. For our particular problem, we consider the simplest case in which the $K$-matrix elements are just constants.

\subsection{Determination of the complete information on the interaction}\label{subsec:Full}
For this part we follow the strategy used in Ref.~\cite{Dai:2023cyo} to determine the compositeness of the $T_{cc}(3875)^+$ state from the mass distribution of Ref.~\cite{LHCb:2021auc} (Fig. 8 of supplementary material).

We start from the potential of Eq.~\eqref{eq:Vij} and the $T$-matrix of Eq.~\eqref{eq:BSeq} and assume that isospin is a good symmetry of the potential (it is a bit broken at the level of the $T$-matrix because of the different masses of the two channels). With the isospin doublets $(D^+, -D^0)$, $(D^{*+}, -D^{*0})$ we write the $I=0,\, 1$ states as
\begin{subequations}\label{eq:isobas}%
\begin{align}
\ket{D^{*}D,I=0}  & = -\frac{1}{\sqrt{2}} \left( \ket{D^{*+}D^0}-\ket{D^{*0}D^+} \right) \,, \\
\ket{D^{*}D,I=1}  & = -\frac{1}{\sqrt{2}} \left( \ket{D^{*+}D^0} + \ket{D^{*0}D^+} \right)\,.
\end{align}
\end{subequations}
Demanding isospin conservation, \textit{i.e.} $\matrixel{I=0}{V}{I=1}=0$, we obtain $V_{22}=V_{11}$. 

In addition, in order to allow for the general case in which the $T_{cc}(3875)^+$ could be a compact tetraquark state, a pure molecular state of the $D^{*+}D^0$ and $D^{*0}D^+$ states, or contain a mixture of both, or other missing two hadron components \cite{Hyodo:2013nka,Aceti:2014ala,MartinezTorres:2014kpc,Montesinos:2023qbx}, we write:
\begin{subequations} \label{eq:Vgen}%
\begin{align}
V_{11} & = V'_{11} + \frac{\alpha}{m_V^2}(s-s_\text{th}) \,, \\
V_{12} & = V'_{12} + \frac{\beta }{m_V^2}(s-s_\text{th}) \,,
\end{align}
\end{subequations}
with $V'_{11}$, $V'_{12}$ constants, and where $s_\text{th}$ is the $D^{*+}D^0$ threshold energy squared,\footnote{In Ref.~\cite{Dai:2023cyo} one takes $s-s_{0}$ instead of $s-s_\text{th}$ in Eq.(\ref{eq:Vgen}) with $s_0$ the $T_{cc}(3875)^+$ mass squared. Since \textit{a priori} $s_0$ is not known from the correlation functions alone, we take $s-s_\text{th}$, but $s-s_\text{th}=s-s_0-(s_\text{th}-s_{0})$, and the constant part $s_\text{th}-s_{0}$ can be reabsorbed in the $V'_{11}$, $V'_{12}$ coefficients, so the two formulations are equivalent.} $\alpha$, $\beta$ are free parameters and $m_V=800\,\text{MeV}$ a typical vector meson mass included to make $\alpha$ and $\beta$ dimensionless.

We should note that the formalism of using Eqs.~\eqref{eq:Vij} and \eqref{eq:BSeq} with the $G$ function of Eq.~\eqref{eq:Gcut} is valid when all the involved particles are stable. In Refs.~\cite{Dai:2023cyo,Feijoo:2021ppq} the $D^*$ width is considered and, as shown in Ref.~\cite{Dai:2022ulk}, this is easily done replacing the $G$ function in Eq.~\eqref{eq:Gcut} by:
\begin{align}\label{eq:GcutDst}
\widehat{G}_i(s) & = \int_{0}^{\qmax} dq\, H_i(q) \frac{1}{\sqrt{s}+\omega_1 + \omega_2} \nonumber \\
&\times  \frac{1}{\sqrt{s}-\omega_1-\omega_2+i\frac{\sqrt{s'}}{2m_{D^*}}\Gamma_{D^*}(s')} ,
\end{align}
with $s'=(\sqrt{s}-\omega_D)^2-{\vec{q}}^{\,\,2}$ and $\Gamma_{D^*}(s')$ as given in Eqs.\,(14) and (15) of Ref.~\cite{Feijoo:2021ppq}. In the limit in which the total width of the vector meson goes to zero, $\widehat{G}_i(s)$ reduces to $G_i(s)$. 

The new formalism has now more parameters: $V'_{11}$, $V'_{12}$, $\alpha$, $\beta$ and $\qmax$ plus the parameter $R$ if it is also unknown. In total six parameters to describe the two correlation functions in the energy range of the experiment. Once these parameters are fixed, one can calculate all properties related to the $T_{cc}(3875)^+$ state: 
 \begin{enumerate}
\item $a_{D^{*+}D^0}$ and $a_{D^{*0}D^+}$.
\item $r_{0,D^{*+}D^0}$ and $r_{0,D^{*0}D^+}$ from the effective range expansion $k \cot\delta(k) \approx -\frac{1}{a}+\frac{1}{2}r_0 k^2$.
\item The existence of a bound state and the binding energy.
\item The couplings $g_{D^{*+}D^0}$, $g_{D^{*0}D^+}$ of the state to the $D^{*+}D^0$ and $D^{*0}D^+$ channels.
\item The width of the state, $\Gamma_{T_{cc}}$.
\item The probabilities $P_1$, $P_2$ of the channels $D^{*+}D^0$, $D^{*0}D^+$ in the wave function of the $T_{cc}(3875)^+$, and $Z=1-P_1-P_2$ the probability of other components than the former ones in the wave function.
\end{enumerate}

In order to evaluate the magnitudes enumerated above, we use the formulas of sections III and IV of Ref.~\cite{Dai:2023cyo}, which we do not reproduce here. The $a_{D^{*+}D^0}$, $a_{D^{*0}D^+}$ scattering lengths are obtained from $T_{11}$ and $T_{22}$ at their respective thresholds and they are complex. They are very similar to those obtained in the limit of $\Gamma_{D^*} \to 0$ except that $a_{D^{*+}D^0}$ is real in such a case, however, $\text{Im}\,a_{D^{*+}D^0}$ with finite $D^*$ width is different from zero but quite smaller than $\text{Re}\,a_{D^{*+}D^0}$. The radii and couplings are evaluated in the limit of $\Gamma_{D^*} \to 0$. Then, the probabilities $P_1$, $P_2$ are given by \cite{Hyodo:2013nka,Aceti:2014ala,MartinezTorres:2014kpc} 
\begin{subequations}\label{eq:Probs}
\begin{align}
P_1 & = -g^2_{D^{*+}D^0}\frac{\partial G_{D^{*+}D^0}}{\partial s} \Big|_{s_0} \,,\\ 
P_2 & = -g^2_{D^{*0}D^+}\frac{\partial G_{D^{*0}D^+}}{\partial s}\Big|_{s_0} \,.
\end{align}
\end{subequations}
If $P_1+P_2$ is unity or very close, this indicates that we have a molecular state made from the $D^{*+}D^0$, $D^{*0}D^+$ channels.

\section{Results}\label{sec:results}

\begin{figure*}[t]\centering
\includegraphics[width=0.85\textwidth,keepaspectratio]{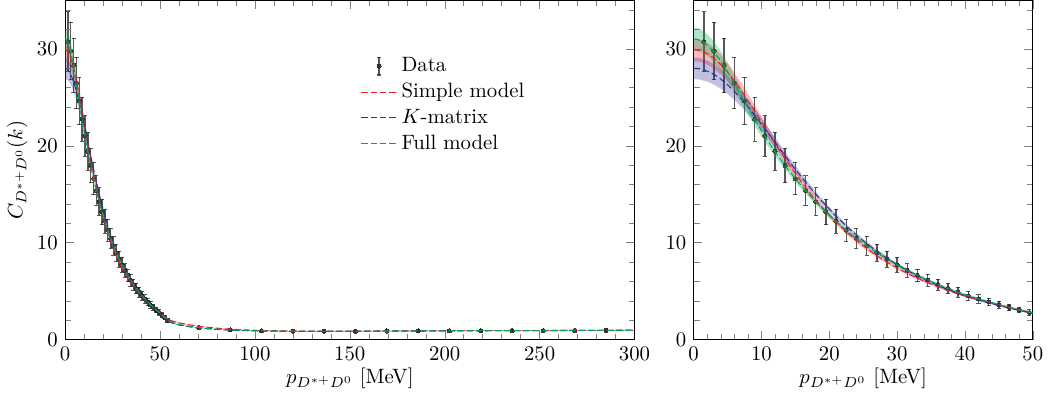}
\includegraphics[width=0.85\textwidth,keepaspectratio]{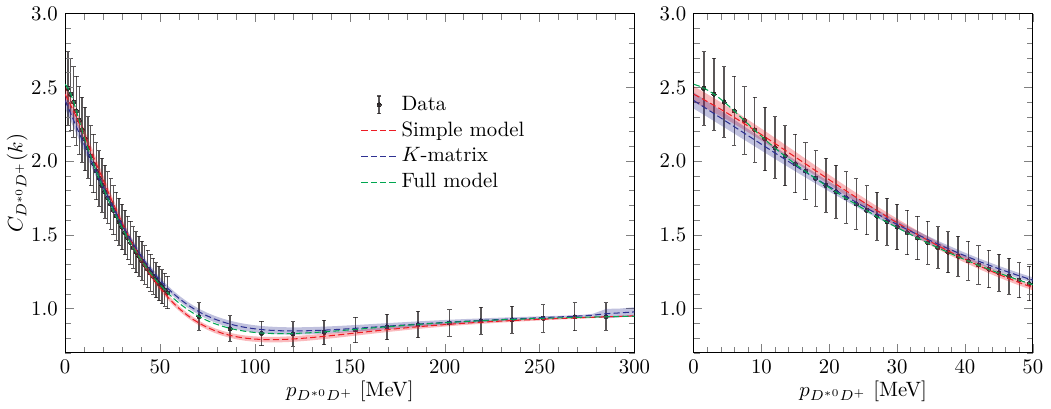}
\caption{Comparison of the synthetic data for the $D^{\ast+}D^0$ (top) and $D^{\ast0}D^+$ (bottom) correlation functions with the results obtained in Subsects.~\ref{subsec:ScLe}, \ref{subsec:KMat}, and \ref{subsec:Full}. The $x$-axis variable is the c.m. momentum of each pair. The right panels show a zoom into the lower momentum region of the corresponding left ones.\label{fig:fits}}
\end{figure*}

To study the inverse problem of obtaining the $T$-matrix from correlation function data, and in the absence of real data for the $D^{\ast+}D^0$--$D^{\ast0}D^+$ system, we generate synthetic data by taking the correlation functions computed in Ref.~\cite{Vidana:2023olz} for $R=1\,\text{fm}$ and picking $N=50$ points in the momentum variable for each channel. We take errors of $10\%$ of the central value as is common in most present experiments. Furthermore, we take more data in the region $k \lesssim 50\,\text{MeV}$, to take into account that this is the region closest to threshold and thus more interesting. The data are shown in Fig.~\ref{fig:fits}, together with the fits that we carry on with the different formalisms described in the previous section. For each formalism, we perform a best fit with MINUIT \cite{James:1975dr,James:1994vla}, and obtain the errors of the parameters (as well as of every other quantity, \textit{e.g.} scattering lengths) by means of the bootstrap technique \cite{Efron:1986hys}. The parameter errors that we obtain in our final fits with the bootstrap method are in good agreement with those derived with MINUIT both with \texttt{MIGRAD} (from the $\chi^2$ hessian) and with \texttt{MINOS}. In what follows, we discuss the different fits. 

\begin{table*}[t]\centering
\begin{tabular}{ccccc} \hline
Model                                              & $a_1\,[\text{fm}]$       & $r_1\,[\text{fm}]$ & $a_2 [\text{fm}]$            & $r_2\,[\text{fm}]$\\ \hline
Subsec.~\ref{subsec:ScLe}                          & $3.85(25)$               & ---                & $2.837( 98) - i 0.972(92)$   & ---                    \\
Subsec.~\ref{subsec:KMat}                          & $8.22(43)$               & ---                & $1.600(170) - i 1.206(62)$   & ---                    \\
Subsec.~\ref{subsec:Full}                          & $7.83(26)(31) - i 1.84(13)(15)$  & $-2.840(47)(280)$       & $2.084( 61)(62) - i 1.219(51)(97)$   & $-0.08(12)(30) - i 2.56(23)(33)$\\  \hline
Exp. \cite{LHCb:2021vvq,LHCb:2021auc}              & $7.10(51) - i 1.85(28)$  & ---                & $1.760(300) - i 1.820(300)$  & ---                      \\
$\Gamma_{D^{\ast}} \to 0$ \cite{LHCb:2021vvq,LHCb:2021auc}              & $6.13(51)$              & $-3.52(50)$        & $1.710(300) - i 1.070(300)$  & $0.26(30) -i 3.27(30)$\\   \hline
\end{tabular}
\caption{Comparison of the $D^{\ast+}D^0$ ($a_1$, $r_1$) and $D^{\ast0}D^+$ ($a_2$, $r_2$) scattering lengths and effective ranges, as obtained with the different methods employed to fit the synthetic correlation function data in Fig.~\ref{fig:fits}, as well as with their experimental determinations. In the last row we also show the results of the experimental analysis model in the limit $\Gamma_{D^\ast} \to 0$, which is the limit in which we compute the effective ranges $r_{1,2}$. For the results of Subsec.~\ref{subsec:Full}, the first (second) uncertainty represent the statistical (systematic) one, computed as described in the text.%
\label{tab:allmethods}%
}%
\end{table*}

\subsection{Scattering length}\label{sec:ResScLe}
We start with the first method described in Subsec.~\ref{subsec:ScLe}. The (real) parameters that we have are  $a_1$, $b_2$, $b'_2$, $\qmax$, and $R$. The best fit has $\chi^2/\text{d.o.f.} =0.08$.\footnote{As a general note, since we have taken relatively large errors, and the synthetic data are generated from a theoretical model and without any real noise, the $\chi^2/\text{d.o.f.}$ that we obtain for all the fits are necessarily going to be small. However, the point that we want to emphasize in this paper is not the goodness of the obtained fits, but the possibility to analyze future, real data with the different amplitudes that we present.}  The parameters are given by:
\begin{subequations}\label{eq:ResScLe}
\begin{align}
a_1   & = 3.85(21)     \, \text{fm}      \,, \\
b_2   & = -0.3149(26)   \, \text{fm}^{-1} \,, \\
b'_2  & = 0.410(56)                      \,, \\
\qmax & = 375(26)       \, \text{MeV}     \,, \\
R     & = 1.038(25)     \, {\rm fm}       \,.
\end{align}
\end{subequations}
The results for the obtained correlation functions are shown in Fig.~\ref{fig:fits}, where the good agreement with the synthetic data can be observed. The scattering length $a_1$ in this method is directly a free parameter, but we can also obtain the scattering length $a_2$ [via Eq.~\eqref{eq:scatl2}], which is shown in Table~\ref{tab:allmethods}. These results should be compared with those extracted from the model employed in Ref.~\cite{LHCb:2021auc} in the analysis of the data in the limit $\Gamma_{D^*} \to 0$,\footnote{We are thankful to Mikhail Mikhasenko for providing us these data.} which are also shown in Table~\ref{tab:allmethods}. As we can see in Table~\ref{tab:allmethods}, the values of $a_{D^{*+}D^0}\,(a_1)$ and $a_{D^{*0}D^+}\,(a_2)$ can be considered only in rough qualitative agreement with the experiment. It is also rewarding to see that the method provides the value of $R$ assumed for the data with great precision.\footnote{The method would also allow for a determination of the $T_{cc}(3875)^+$ binding energy, for instance looking for a pole in the $T_{11}$ amplitude. This would give a well-known result for the binding energy, $B = 1/(2\mu a_1^2)$, with $\mu$ the $D^{\ast+}D^0$ reduced mass. The result is $B \simeq 1.35\,\text{MeV}$, larger than the experimental one \cite{LHCb:2021vvq,LHCb:2021auc}. However, since the goal of the method was to obtain an estimation of the scattering lengths and we employ below more sophisticated methods, we do not discuss in detail this result.}

\subsection{\boldmath Results with the $K$-matrix formalism}

We now fit the synthetic data employing the $T$-matrix described in Subsec.~\ref{subsec:KMat}, taking as free (and real) parameters $K_{11}$, $K_{12}$, $K_{22}$, $\qmax$, and $R$. The fit has $\chi^2/\text{d.o.f.} = 0.09$, and the comparison between the obtained correlation functions with the synthetic data is shown in Fig.~\ref{fig:fits}. For the parameters we obtain the values:
\begin{subequations}
\begin{align}
K_{11} & = \hphantom{-}1368(110)\,,\\
K_{22} & = \hphantom{-}1227(40)\,,\\
K_{12} & =           - 1308(55)\,,\\
R & = 0.97(4)\,\text{fm},\\
\qmax & = 342(38)\,\text{MeV}\,.
\end{align}
\end{subequations}
We note that $K_{11} \simeq K_{22}$, as expected from isospin symmetry. On the other hand, the cutoff is around $2\sigma$ with respect to the original value, $\qmax = 418\,\text{MeV}$. The model of Subsec.~\ref{subsec:KMat} for the $T$-matrix is different from that used to generate the synthetic data, and, furthermore, the former does not depend on the cutoff. Therefore, the dependence of $\widetilde{G}$ on $\qmax$ tends to compensate this effect. Since the dependence is mild, because the $j_0(qr)$ in $\widetilde{G}$ already acts as a regulator inEq.~\eqref{eq:GtildeOriginal}, this somewhat large deviation of $\qmax$ is reasonable.

\begin{table}[t]\centering
\begin{tabular}{ccc} \hline
Model                                  & $B\,[\text{keV}]$ & $\Gamma\,[\text{keV}]$ \\ \hline
Subsec.~\ref{subsec:KMat}              & $218(17)$         & ---      \\
Subsec.~\ref{subsec:Full}              & $341(16)(11)$     & $38(2)(1)$ \\
Exp. \cite{LHCb:2021vvq,LHCb:2021auc}  & $360(40)$         & $48(2)$ \\ \hline
\end{tabular}
\caption{Comparison of the $T_{cc}(3875)^+$ binding energy ($B$) and width ($\Gamma$), as obtained with the different methods employed to fit the synthetic correlation function data in Fig.~\ref{fig:fits}, as well as with their experimental determinations. For the results of Subsec.~\ref{subsec:Full}, the first (second) uncertainty represent the statistical (systematic) one, computed as described in the text.\label{tab:allmethodsBindingWidth}}
\end{table}

We can also compute the scattering lengths and the $T_{cc}(3875)^+$ binding energy, which are shown in Tables~\ref{tab:allmethods} and \ref{tab:allmethodsBindingWidth}, respectively. As can be seen, there is a clear improvement with respect to the previous method in the scattering length determinations, which are now in better agreement with the experimental ones, in particular $a_{1}$. The binding energy is also close to the experimental one, although around $3\sigma$ away from it. 

One could think of improving the parameterization by including higher-order ($p^2$) terms in the $K$-matrix elements and/or imposing isospin symmetry (forcing $K_{11} = K_{22}$). These modifications would lead to parameterizations closer to the full model and, since we employ the latter in the next section, we refrain here from doing such an exercise.

\subsection{Results with the general method}

We discuss now the results obtained with the full method, described in Subsec.~\ref{subsec:Full}. In a first step, we consider as free parameters $V'_{11}$, $V'_{12}$, $\alpha$, $\beta$, $\qmax$, and $R$.\footnote{We find it more advantageous to work with the combinations $V'_{11} \pm V'_{12}$ and $\alpha \pm \beta$, which correspond to the isospin-definite combinations, as shown below.} From the fits obtained with this set of free parameters, two issues can be observed. First, the parameters $\alpha$ and $\beta$ are difficult to constrain, and furthermore, they are compatible with zero.\footnote{For reference, in the best fit we find the values $\alpha - \beta =-60(340)$ and $\alpha + \beta = -70(2700)$, with errors given by \texttt{MIGRAD}, and even larger errors from the \texttt{MINOS} algorithm. Note that the relevant combination here is $\alpha - \beta$, since this would be the isospin zero combination.} Therefore, without any loss of generality, we can neglect them. Second, the cutoff $\qmax$ is largely correlated (with a correlation coefficient $r \simeq 0.99$) with the parameters $V'_{11}$ and $V'_{12}$ and, therefore, they cannot be simultaneously fitted in a proper manner. This correlation is of course well known, and it corresponds to the fact that, just taking a single $I=0$ channel for simplicity, one would have $T^{-1}=(V_{11}-V_{12})^{-1}-G$, and there is always the possibility of a trade off between $V_{11}-V_{12}$ and $\qmax$, such that choosing $\delta (V_{11}-V_{12})= \delta G$ at the pole one does not change its position. Then, the amplitude close to the pole position (which in turn is close to the threshold) barely changes. The dependence on $\qmax$ appears not only in the $T$-matrix, as just discussed, but also in the $\widetilde{G}_i(s,r)$ functions. However, the dependence of the latter on the cutoff is mild, as shown in Refs.~\cite{Vidana:2023olz,Albaladejo:2023pzq}.

As a consequence of the previous considerations, in a second step we set $\alpha = \beta = 0$, and take a fixed value for the cutoff, namely $\qmax = 418\,\text{MeV}$ (which is the value used in Ref.~\cite{Feijoo:2021ppq} to compute the $T$-matrix, from which we have generated the synthetic data), such that the free parameters are now $V'_{11}$, $V'_{12}$, and $R$. This final set of parameters are fitted to the synthetic data, and the results of the fit are shown in Fig.~\ref{fig:fits}, compared with the data and the results obtained with the rest of the formalisms. The best fit has $\chi^2/\text{d.o.f.} \sim 10^{-4}$, since this is a model for the $T$-matrix almost identical to the one employed to generate the synthetic data. The parameter values are:
\begin{subequations}\label{eq:fitFullparameters}
\begin{align}
V_{I=0} & \equiv V'_{11} - V'_{12} = -435.88(88)\,~, \label{eq:V0FullMethod}\\
V_{I=1} & \equiv V'_{11} + V'_{12} = \hphantom{-} 600(330)\,~, \label{eq:V1FullMethod}\\
R & = 1.008(24)(11)\, \text{fm}\,~, \label{eq:R1FullMethod}
\end{align}
\end{subequations}
with a correlation matrix:
\begin{equation}\label{eq:correlation}
\begin{pmatrix}
1 & -0.65 & \hphantom{-}0.95 \\
  &    1  & -0.70 \\
  &       &    1
\end{pmatrix}~,
\end{equation}
being rows and columns ordered as in Eq.~\eqref{eq:fitFullparameters}. 

As discussed earlier, we have taken a fixed value of the cutoff $\Lambda$, but we nonetheless take into consideration the uncertainty induced by the value of $\Lambda$ in the following manner. We perform fits of the free parameters to the synthetic data for values of $\Lambda$ in the range $\left[ 350,500 \right]\,\text{MeV}$. We obtain the maximum and the minimum value of every quantity, such as the scattering lengths or the $T_{cc}(3875)^+$ binding energy, and we take half the difference between both values as an estimation of a systematic uncertainty. Thus, in this section, whenever two uncertainties are given [for instance in Eq.~\eqref{eq:R1FullMethod}], the first one refers to the statistical uncertainty obtained with the bootstrap method, and the second one the systematic uncertainty from the cutoff variation just described. This systematic uncertainty is not shown for $V_{I=0,1}$ [Eq.~\eqref{eq:V0FullMethod} and \eqref{eq:V1FullMethod}] because, as has been also discussed, the correlation of these parameters with $\Lambda$ is almost one. Therefore, we just quote the fitted values that we obtain for $\Lambda=350\,\text{MeV}$, $V_{I=0} = -532$ and $V_{I=1}=300$, and for $\Lambda=500\,\text{MeV}$, $V_{I=0} = -355$ and $V_{I=1}=-2001$. As said, the spread here is much larger than the statistical uncertainty.

As can be seen [Eqs.~\eqref{eq:V0FullMethod} and \eqref{eq:V1FullMethod}], the isoscalar matrix element $V_{I=0}$ is much more constrained than its isovector counterpart $V_{I=1}$. This is due to the isoscalar nature of the $T_{cc}(3875)^+$ state. The pole of this state, being below and very close to the threshold, enhances the correlation function near the $D^{\ast+}D^0$ threshold, such that $C_{D^{\ast+}D^0}(0) \simeq 30$. We note that, as discussed right after Eq.~\eqref{eq:source}, we are able to recover in Eq.~\eqref{eq:R1FullMethod} the original value $R=1\,\text{fm}$ for the size of the source function with which the synthetic data were generated. The large correlation between $R$ and $V_{I=0}$ in Eq.~\eqref{eq:correlation} can be easily understood. In Ref.~\cite{Albaladejo:2023pzq} it was shown that, for a single channel, in the Lednicky-Lyuboshits approximation \cite{Lednicky:1981su}, one obtains that the deviation from 1 of $C(k=0)$ depends on the ratio $a/R$, with $a$ the scattering length, which is determined from the matrix element $V_{I=0}$. In a system different from $D^{\ast}D$, with a smaller scattering length or, equivalently, without the prominent enhancement induced by the $T_{cc}(3875)^+$, we would expect this correlation to be smaller. Furthermore, we have tested that using \texttt{MINOS}, which estimates the (negative and positive) errors of a parameter by varying it and fitting the other free parameters until the $\chi^2$ deviates from the minimum in one unit, we get uncertainties very similar to the statistical one shown in Eq.~\eqref{eq:R1FullMethod}, obtained from the bootstrap method. Likewise, the (parabolic) error obtained from the $\chi^2$ hessian (with \texttt{MIGRAD}) is very similar, too. Therefore, despite the large correlation coefficient between $R$ and $V_{I=0}$ in Eq.~\eqref{eq:correlation}, the method proposed here allows one to determine $R$ with about $2$--$3\%$ of statistical uncertainty. Hence, we expect that this will also be the case with real data.

With this $T$-matrix parameterization we can also compute the scattering lengths and effective ranges,\footnote{The effective range for both channels are computed in the limit $\Gamma_{D^{\ast}} \to 0$, to ensure that the $ik$ term in the $T$-matrix elements is properly removed for the effective range expansion of $k\cot\delta(k)$.} shown in Table~\ref{tab:allmethods}, as well as the $T_{cc}(3875)^+$ binding energy and its width, shown in Table~\ref{tab:allmethodsBindingWidth}. It can be seen that there is quite good agreement with the experimental determination \cite{LHCb:2021vvq,LHCb:2021auc} of the same quantities, also shown in Tables~\ref{tab:allmethods} and \ref{tab:allmethodsBindingWidth}. We also note that the method allows for a determination of the uncertainty of these quantities similar or smaller than the experimental ones. Shall the $D^{\ast+}D^0$ and $D^{\ast0}D^+$ correlation functions be measured with an uncertainty of around $10\%$, as we have imposed in our synthetic data, the $D^{\ast+}D^0$ and $D^{\ast0}D^+$ interactions, as well as the $T_{cc}(3875)^+$ parameters, could be greatly constrained.

We can also compute the $T_{cc}(3875)^+$ couplings to the two channels, obtaining:
\begin{subequations}\label{eq:couplings}
\begin{align}
g_1 & = \hphantom{-}3853(36)(33)\,\text{MeV}\,,\\
g_2 & =           - 4131(73)(35)\,\text{MeV}\,,
\end{align}
\end{subequations}
which in turn allow for the determination of the $D^{\ast+}D^0$ and $D^{\ast0}D^+$ probabilities,
\begin{subequations}\label{eq:probabilities}
\begin{align}
P_1 & = - g_1^2 \left. \frac{\partial G_1}{\partial s} \right\rvert_{s=s_0} = 0.691(9)(15)\,,\\
P_2 & = - g_2^2 \left. \frac{\partial G_2}{\partial s} \right\rvert_{s=s_0} = 0.309(9)(15)\,.
\end{align}
\end{subequations}
The values obtained above fulfill $P_1 + P_2 = 1$ by construction, indicating a fully molecular state \cite{Gamermann:2009uq}, since we have neglected $\alpha$ and $\beta$. We recall here that in our first step of the analysis we obtained $\alpha$ and $\beta$ values largely compatible with zero. This result, in turn, is an unavoidable consequence given the fact that the synthetic data are generated with a model \cite{Feijoo:2021ppq} for the $T$-matrix in which the $T_{cc}(3875)^+$ state turns out to be almost purely molecular (similar results are obtained in Refs.~\cite{Albaladejo:2021vln,Du:2021zzh,Albaladejo:2022sux}). However, this is not the point that we want to emphasize here, rather, the possibility of obtaining these quantities with relative fine precision, say of a few percent, shall these correlation functions be measured. This proves the ability of the method to extract the nature of states from the analysis of correlation data with great precision.

We would like to mention here that in the first step of the calculations where $\alpha$ and $\beta$ are free parameters, the results that we obtain for the observables are very similar to and compatible within errors with those in Table~\ref{tab:allmethods}. In the case where $\alpha$ and $\beta$ are free, $P_1+P_2$ is not exactly $1$ and one obtains $P_1 + P_2 = 1.002(49)$, which is indeed compatible with $P_1 + P_2 = 1$, and the uncertainty is a little bigger than the sum in quadratures of the errors of $P_1$ and $P_2$ in Eqs.~\eqref{eq:probabilities}.

\section{Conclusions}
In this work we investigated the potential of the inverse problem in the femtoscopic correlation functions, in which starting from the correlation functions of some related channels we determine the interaction between these channels. The data are analyzed by means of a coupled channel unitary scheme which has the freedom to accommodate missing channels relevant to the interaction, as well as components of some genuine states of non-molecular nature. This leaves a few parameters to be determined from the data, together with the parameter that gives the source size. We applied the method to pseudo-data of the $D^{*+}D^0$ and $D^{*0}D^+$ correlation functions and extracted from there all the possible information. The results obtained are remarkable and we found that using these data, all above the threshold of the two channels, we could determine that there was a bound state of the system, that this state was of molecular nature, with negligible room for a non-molecular state. We determined the scattering length and the effective range for the two channels, the probabilities to have each of the channels in the wave function of the bound state and also, remarkably, the size of the source parameter, $R$, all of them with a good precision and in agreement with the experimental information used to constrain the $T$-matrix employed in the generation of the synthetic data (correlation functions).

In order to better appreciate the precision in the magnitudes obtained with the general method, we also used two other different methods, one very simple which does not require to use any intermediate interaction potentials between the channels, and which provides the scattering amplitudes at a qualitative level. Another one, based on the common $K$-matrix approach, provides much better results for the scattering amplitudes, but however does not allow to obtain the molecular probabilities which rely upon $\partial G/\partial s$, and $\text{Re}\,G$ is not fully accounted for in the $K$-matrix approach.

In conlusion, the full method proposed here is ideal to analyze correlation functions and extract from them valuable information on hadron interactions, which will be useful in the future to study the interaction of hadron states which cannot be formed otherwise, as well as to determine the nature of bound states or resonances.

\begin{acknowledgments}
This work was supported by the Spanish Ministerio de Ciencia e Innovaci\'on (MICINN) and European FEDER funds under Contracts No.\,PID2020-112777GB-I00, and by Generalitat Valenciana under contract PROMETEO/2020/023. This project has received funding from the European Union Horizon 2020 research and innovation programme under the program H2020-INFRAIA-2018-1, grant agreement No.\,824093 of the STRONG-2020 project. M.\,A. and  A.\,F. are supported through Generalitat Valencia (GVA) Grants Nos.\,CIDEGENT/2020/002 and APOSTD-2021-112, respectively. M.\,A. and A.\,F. thank the warm support of ACVJLI. 
\end{acknowledgments}

\bibliographystyle{apsrev4-1}
\bibliography{References.bib}

\end{document}